\documentclass[twocolumn,showpacs,amsmath,amssymb,prl,superscriptaddress]{revtex4}
\usepackage{graphicx}
\usepackage{bm}
\usepackage{dcolumn}

\begin{document}
\title{Kekul\'e textures, pseudo-spin one Dirac cones and quadratic band crossings in a graphene-hexagonal indium chalcogenide bilayer.}

\author{Gianluca Giovannetti}
\affiliation{CNR-IOM-Democritos National Simulation Centre and International School
for Advanced Studies (SISSA), Via Bonomea 265, I-34136, Trieste, Italy}
\affiliation{Institute for Theoretical Solid State Physics, IFW-Dresden, PF 270116, 01171 Dresden, Germany}
\author{Massimo Capone}
\affiliation{CNR-IOM-Democritos National Simulation Centre and International School
for Advanced Studies (SISSA), Via Bonomea 265, I-34136, Trieste, Italy}
\author{Jeroen van den Brink}
\affiliation{Institute for Theoretical Solid State Physics, IFW-Dresden, PF 270116, 01171 Dresden, Germany}
\author{Carmine Ortix}
\affiliation{Institute for Theoretical Solid State Physics, IFW-Dresden, PF 270116, 01171 Dresden, Germany}

\begin{abstract}
Using density-functional theory, we calculate the electronic bandstructure of single-layer graphene on top of  hexagonal In$_{2}$Te$_{2}$ monolayers. The geometric configuration with In and Te atoms at centers of carbon hexagons leads to a Kekul\'e texture with an ensuing bandgap of 20 meV. The alternative structure, nearly degenerate in energy, with the In and Te atoms on top of carbon sites is characterized instead by gapless spectrum with the original Dirac cones of graphene reshaped,  depending on the graphene-indium chalcogenide distance, either in the form of an undoubled pseudo-spin one Dirac cone or in a quadratic band crossing point at the Fermi level. These electronic phases harbor charge fractionalization and topological Mott insulating states of matter.
\end{abstract}

\maketitle

Since the pioneering work of Esaki and Tsu back in 1970 \cite{EsakiTsu}, superlattices have attracted a huge interest because mostly because they can be used to tailor the electronic properties of conventional materials. 
Two alternative methods to create superlattices of two-dimensional (2D) electrons have been devised. The first method is based on the lithographic patterning of semiconductor surfaces \cite{Albrecht}. The alternative method, which has received a boost with the experimental discovery of graphene \cite{Geim1} and other two-dimensional (2D)  nanomaterials such as germanene and silicene, relies on the existence of long-period moir\'e patterns formed by two slightly incommensurate 2D lattices with same crystal symmetry placed on top of each other. 
Graphene on hexagonal boron nitride (hBN) \cite{GiovannettiGRBN}  has emerged as a remarkable example of such a superstructure \cite{OrtixInc,Wallbank}. This, has led to 
a plethora of phenomena including cloning of Dirac fermions\cite{Yankowitz,Geim2} and the experimental realization of the Hofstadter butterfly \cite{KIM,Hunt}  -- a fractal spectrum theoretically predicted in conventional 2D electron gases subject to external ultralarge magnetic fields\cite{Hofstadter}.
As recently shown for graphene on hBN \cite{Woods}, large moir\'e superstructures can  undergo a structural transition to a commensurate state. This occurs whenever the period of the superstructure is so large that it becomes energetically favorable to adjust the two lattices to become commensurate, loosing in elastic energy but gaining in van der Waals energy.

It is then conceivable that qualitatively different short-period commensurate graphene superlattices can be engineered using as substrates the large number of semiconductors with an hexagonal surface layer, layered hexagonal crystals, and the corresponding 2D crystals that can be produced thereof with larger unit cells. The simplest example of a commensurate graphene superstructure, for instance, can be manufactured by placing graphene on top of hexagonal underlays with a unit cell almost three times bigger than the graphene one.  These include three-dimensional materials -- PtTe$_2$, h-GaTe, InAs, CdSe \cite{Falko} to name but a few -- but also next-generation 2D crystals such as gallium and indium chalcogenides (Ga,In)$_2$X$_2$ with X$=$S, Se, Te \cite{Falko2,Falko3}. 

In a commensurate $\sqrt{3} \times \sqrt{3}$ graphene superlattice, intervalley scattering of the graphene Dirac electrons can potentially lead to several phenomena including the opening of a gap by means of a Kekul\'e distortion, realized so far only in molecular graphene \cite{Gomes}.
By considering, with a set of density-functional theory (DFT) calculations, a prototypical  commensurate graphene superlattice with tripled unit cell, namely graphene  on top of  hexagonal In$_{2}$Te$_{2}$ monolayers, in this Letter we demonstrate the opening of a  Kekul\'e bandgap of $\simeq 20$ meV -- an energy almost as large as $k_BT$ at room temperature. 
Domains of the Kekul\'e phase generate the presence of a stable fermionic zero mode with fractional charge \cite{Hou} obeying fractional exchange statistics \cite{Sera,Ryu}-- the basic building 
block of a topological quantum computer. 
For an alternative, almost degenerate, geometric configuration, the Dirac cones of pristine graphene are instead reshaped either in the form of a pseudo-spin one Dirac cone \cite{Green} or in a quadratic band crossing point at the Fermi level, which is prone to many-body instabilities towards non-trivial topological states of matter \cite{OrtixArX}.

Indium chalcogenides 
take several forms including tetragonal, rhombohedral, cubic, monoclinic and orthorhombic phases as well as an hexagonal structure. While indium selenide exists in this form in nature, indium sulfide and indium telluride exhibit orthorhombic and tetragonal structures respectively, but this does not preclude the possibility to create metastable structures in the hexagonal phase. This layered structure consists of two vertically aligned hexagonal sublayers of indium atoms sandwiched between two vertically aligned hexagonal sublayers of chalcogen atoms (X) \cite{Falko2}. Viewed from above, a In$_2$X$_2$ monolayer forms a two-dimensional honeycomb lattice. The optical and electronic properties of these single-layer crystals show that they are semiconductors with an indirect bandgap of $\sim$ 1.5 eV \cite{Falko2}. From here onwards, we will restrict to consider indium telluride monolayers since its geometry is such that the distances between In and Te sites are $2.73 \AA$ and $5.50 \AA$ respectively, which gives a planar lattice constant of $4.23 \AA$. This, in turn, gives a lattice mismatch less than $1 \%$ between the indium chalcogenide monolayer and a perfectly commensurate $\sqrt{3} \times \sqrt{3}$ graphene superlattice (with superlattice constant of $4.26 \AA$).  Since this lattice mismatch is even smaller than the $1.8 \%$ percent lattice mismatch between graphene and hBN, it is reasonable to assume that in a graphene-In$_2$Te$_2$ bilayer, the carbon atoms will stretch to follow the periodicity of the indium chalcogenide monolayer, at least for small misorientation angles between the two 2D crystals. 

\begin{figure}
\includegraphics[width=.99\columnwidth,angle=-0]{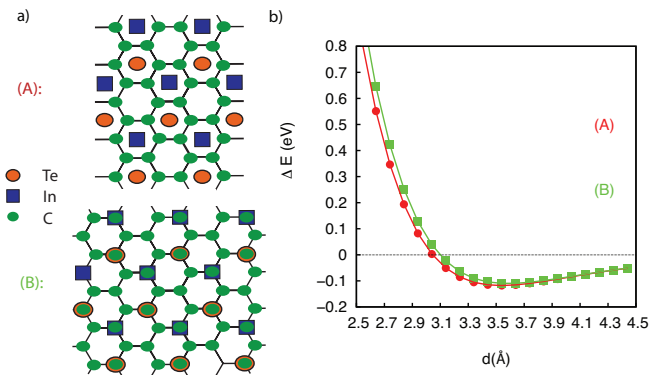}
\caption{(Color online) (a) Top view of the two geometrical configurations (A) and (B) of graphene and In$_2$Te$_2$ layers. (b) Binding energy $\Delta$E of for the two (A) and (B) configurations as a function of the distance $d$ between the graphene sheet and the In$_2$Te$_2$ monolayer.}
\label{fig1}
\end{figure}

We consider two inequivalent configurations of a perfectly commensurate graphene-In$_2$Te$_2$ bilayer, related to each other by a translation of the graphene sheet with respect to the In$_2$Te$_2$ monolayer, see Fig. \ref{fig1}(a): 

(1) the (A) configuration with the In and Te atoms sitting at the center of carbon hexagons; and 

(2) the (B) configuration with the In and Te atoms sitting on top of carbon sites. 

The self-consistent calculations based on DFT\cite{DFT} have been performed with the Vienna ab initio simulation package VASP \cite{VASP} using a plane-wave basis and a kinetic energy cutoff of 500 eV. The Brillouin zone summations have been carried out with the tetrahedron method and a 24$\times$24$\times$1 grid in the Brillouin zone (BZ). We have included a dipole correction to avoid interactions between periodic images along the z direction \cite{DipoleCorr} and represent the vacuum above the graphene sheet with an empty space of 12 $\AA$.
We have performed the  same calculations using the Quantum Espresso package \cite{QE} and found an excellent agreement.

In Fig.~\ref{fig1}(b) we show the total energies of the two configurations (A) and (B) as a function of the distance $d$ between the In$_2$Te$_2$ monolayer and the graphene sheet using the local density approximation (LDA) to the exchange correlation potential \cite{LDA}. The equilibrium separation for configuration (A) of $\sim$ 3.5 $\AA$ is slightly smaller than configuration (B). For all distances, the two configurations are very close in energy within the accuracy of our calculations, although we find that configuration (A) has a total energy a few meV lower than configuration (B). 

Having established the near energetic degeneracy of the two configurations, we have then computed the corresponding electronic structure for both configuration (A) and configuration (B). 
\begin{figure}
\includegraphics[width=.65\columnwidth,angle=-90]{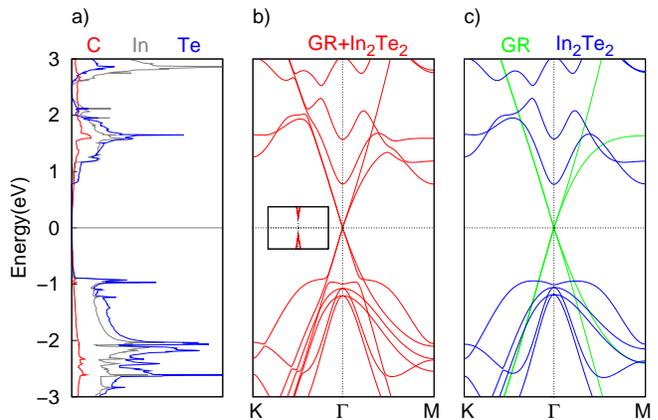}
\caption{(Color online) (a) Projected density of states with C, In, and Te character (b) Band structure of the graphene-In$_2$Te$_2$ bilayer for configuration (A) at the equilibrium distance. (c) Band structure of isolated graphene and In$_2$Te$_2$ single layers.}
\label{fig2}
\end{figure}
In Fig. \ref{fig3}(a)  we show the projected density of states (DOS) and the band structure of our graphene-In$_2$Te$_2$ heterostructure assuming the (A) configuration at the equilibrium distance.
For the In$_2$Te$_2$-derived bands, we find a well defined gap nearly identical to the LDA gap value found for In$_2$Te$_2$ monolayers \cite{Falko2}. 
Within this indium chalcogenide gap, the bands have a predominant carbon character suggesting a weak hybridization as explicitly demonstrated from the comparison of the band structures of the graphene-In$_2$Te$_{2}$ bilayer [c.f. Fig.~\ref{fig3}(b)] and that of graphene and In$_2$Te$_{2}$ as isolated single layers [c.f. Fig.~\ref{fig3}(c)]. On the eV scale of Fig.~\ref{fig3}(a) the original Dirac cones, which are folded at the $\Gamma$ point of the BZ due to the tripling of the graphene unit cell, appears to be preserved [c.f. Fig.~\ref{fig3}(b)]. However, a closer inspection around that point in the BZ reveals that a gap of 16 meV is opened and the dispersion is quadratic. 

It is well-known that within the LDA approximation the exchange correlation potential is not accurately described, resulting in an underestimation of the band gap precisely as for the case of graphene on hBN \cite{GiovannettiGRBN,GiovannettiDopingGR,Menno,Sachs}. To overcome this problem, we have thus performed calculations based the Heyd-Scuseira-Ernzerhof (HSE) hybrid functional\cite{HSE}. We have opted for HSE since it is much faster than other many-body perturbation methods, and provides performances well in agreement with G$_0$W$_0$ or GW calculations \cite{Fuchs}.
The HSE calculations reveal an increase of the bandgap to 20 meV -- a value corresponding almost to $k_{B} T$ at room temperature.

\begin{figure}
\includegraphics[width=.99\columnwidth,angle=-0]{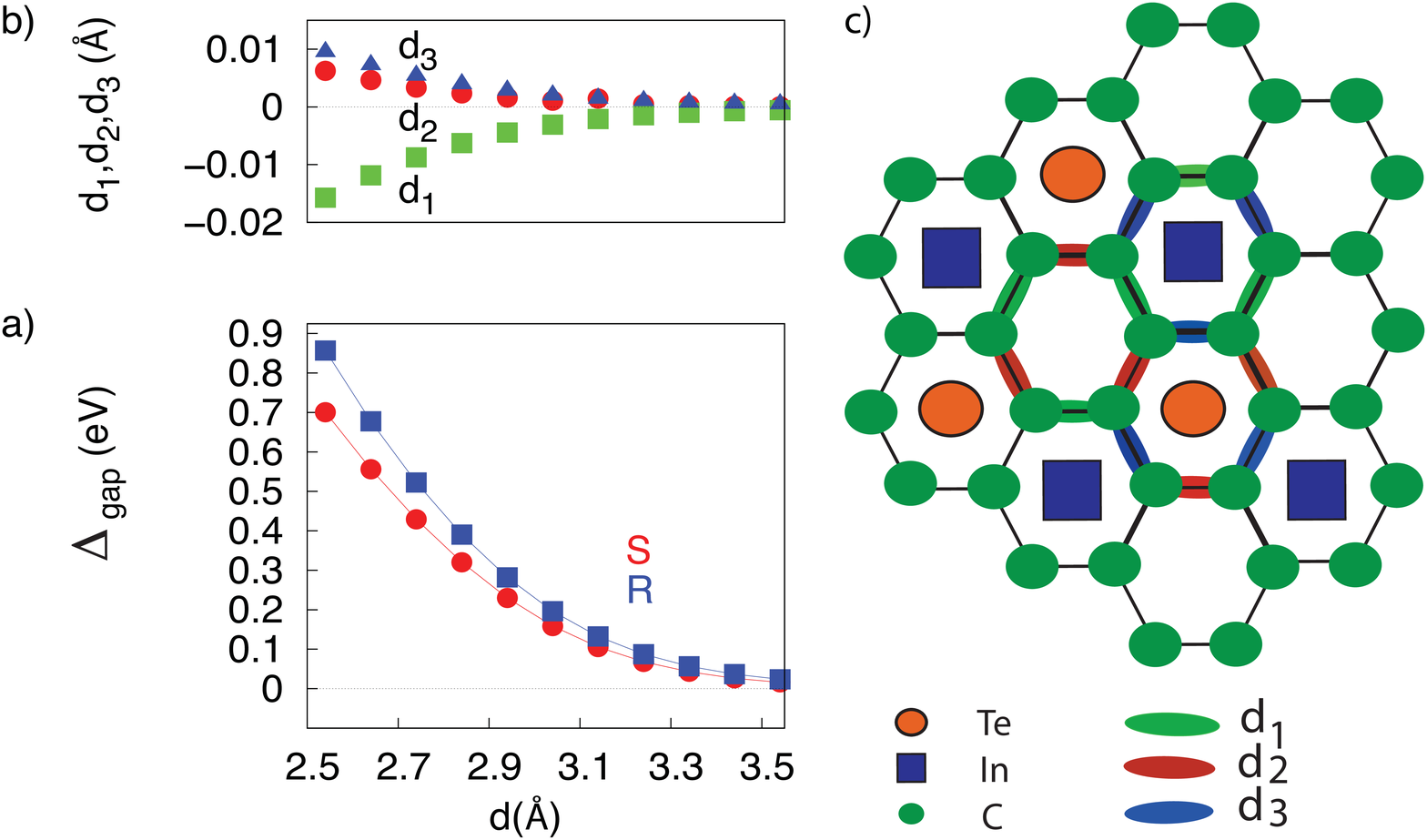}
\caption{(Color online) (a) Values of the band gap as function of the distance $d$ between graphene and In$_2$Te$_2$  layers for unrelaxed (S) and relaxed (R) C atomic coordinates (b) Values of the three inequivalent C-C (d$_1$, d$_2$, d$_3$) bonds as a function of the distance $d$. (c) Schematic view of the three inequivalent C-C (d$_1$, d$_2$, d$_3$) bonds for configuration (A). All the data are obtained within LDA}
\label{fig3}
\end{figure}

To investigate the stability of the bandgap opening mechanism in our graphene-based heterostructure, we computed the electronic structure as a function of the distance $d$ between the two 2D crystals, resorting to the computationally cheaper LDA. We find that the presence of the bandgap at the $\Gamma$ point in the BZ is robust and the gap value increases up to  $0.7\, $eV for  $d \simeq 2.5 \AA$ [c.f. circles (S) in Fig.~\ref{fig3}(a)]. 
The nature of this bandgap opening mechanism is very different from the sublattice symmetry-breaking mechanism expected for instance in graphene on hBN \cite{GiovannettiGRBN}. The occurrence of massive Dirac fermions  in the present heterostructure is indeed due to the appearance of a substrate-induced Kekul\'e phase. 
To explicitly prove this point, we have relaxed the planar positions of the carbon atoms at each distance $d$. In the ensuing stable configuration, the carbon atoms have three different sets of  distances (d$_1$, d$_2$ and d$_3$) whose behavior, as a function of the distance between the graphene sheet and the In$_2$Te$_{2}$ monolayer, is shown in Fig. \ref{fig3}(b). 
These distortions render a Kekul\'e texture, schematically shown in Fig.~\ref{fig3}(c), entirely due to the asymmetry of the In$_2$Te$_2$ ionic potentials. The Te and In triangular planes are indeed at different distances from the graphene layer and this stacking influences the effective potential acting on the carbon-carbon bonds. Te ions are closer to the graphene  layer and lead to a decrease in the carbon-carbon bonds aligned with the corresponding Bravais lattice vectors  [c.f. d$_1$ in Fig. \ref{fig3}(c)]. The remaining carbon-carbon bonds [c.f. d$_2$, d$_3$ in Fig. \ref{fig3}(c)] instead increase due to the larger distance of the indium planes from the graphene layer. The fact that the values of the band gap are further increased upon the ionic relaxation [c.f. squares (R) in  Fig.~\ref{fig3}(a)] confirms that the opening of the gap is related entirely to the inequivalence of the carbon-carbon bonds in the graphene layer. 

 The Kekul\'e phase is topologically trivial but it is characterized by a ${\cal Z}_3$ order parameter corresponding to the three degenerate Kekul\'e ground states obtained by translating the patterns shown in Fig.~\ref{fig3}(c) by the graphene Bravais primitive lattice vectors. Single quantized vortices in the phase of the order parameter, which can be realized in practice at a Y junction of three inequivalent Kekul\'e domains \cite{Bergam} have been shown to possess a stable fermionic zero mode, which carries fractional charge $\pm e/2$ \cite{Hou} and is characterized by fractional exchange statistics \cite{Sera,Ryu}. The value of the Kekul\'e gap in graphene-In$_2$Te$_{2}$ bilayers is large enough to ensure a possible use of this heterostructure for devices potentially relevant for topological quantum computation. We emphasize that the onset of the Kekul\'e phase can be realized not only in the present graphene-In$_2$Te$_2$ bilayers but also using as substrates the large number of materials almost commensurate with $\sqrt{3} \times \sqrt{3}$ graphene superlattice, which share the same structural symmetry. 

\begin{figure}
\includegraphics[width=.65\columnwidth,angle=-90]{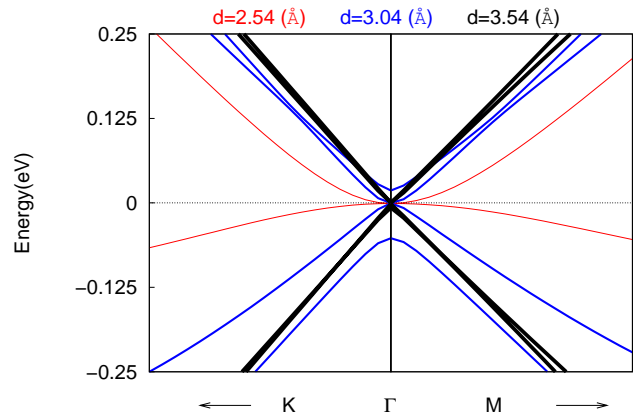}
\caption{(Color online) Bandstructure of graphene-In$_2$Te$_2$ bilayer in configuration (B) along high symmetry directions at some selected layer distances.}
\label{fig4}
\end{figure}

Next, we discuss the electronic bandstructure in the nearly degenerate configuration (B). Fig.~\ref{fig4} shows the band structure at selected distances between the graphene layer and the In$_2$Te$_2$ single layer along high-symmetry directions in the BZ. Zooming in on the $\Gamma$ point, we find the absence of a full bandgap. The original Dirac cones do not acquire any finite mass but  are rather reshaped in a qualitatively different form, which is strongly susceptible to the graphene- In$_2$Te$_2$ layer distance. At the LDA equilibrium distance $d \sim 3.54$ $\AA$,  the band structure  shows a single massless Dirac cone close to the $\Gamma$ point with two additional spin-degenerate bands touching the cone at the Dirac point. This particular form of the electronic structure close to the Fermi level has been dubbed as pseudo-spin one Dirac cone \cite{OrtixArX} since it represents the time-reversal analog of the spin-one Dirac cone predicted in microscopic models of the Kagome lattice with staggered magnetic flux \cite{Green}. In close proximity to the Fermi level, indeed, an effective Hamiltonian for the corresponding conical-like spectrum would be of the form ${\cal H} \propto {\bf k} \cdot {\bf L}$ where the ${\bf L}$ operators generate the three-dimensional pseudo-spin one representation of ${\cal SU}(2)$. 

The conical form of the electronic structure at the $\Gamma$ point of the BZ is lost as one moves away from the LDA optimal equilibrium distance. Considering both larger and smaller distances $d=2.54, 3.04$ $\AA$ the low-energy dispersion is again gapless but a quadratic band crossing point (QBCP) occurs at the Fermi level. QBCP's in two-dimensional systems have recently attracted considerable attention since they are characterized by a finite DOS but do not have a Fermi surface. Electronic interactions are marginally relevant in the renormalization group sense \cite{Sun,Dora}, leading to weak coupling instabilities towards nematic and topological Mott insulating states:  quantum states of matter with an insulating bulk gap generated by interactions and topologically protected edge states. 
The appearance of both a pseudo-spin one Dirac cone and a QBCP can be rationalized by considering again the effect of the ionic potentials of the In$_2$Te$_2$ monolayers on the graphene layers. The In and Te atoms sits indeed on top of two carbon sites belonging to the same original graphene sublattice, giving rise to strongly localized, yet different ionic potentials. 
These different on-site energies realize, in turn, a sublattice symmetry breaking mass for the Dirac fermions as well as a specific combination of constant non-Abelian ${\cal SU}(2)$ gauge fields \cite{OrtixArX}, which give rise to a non-Abelian magnetic field similar to the one predicted to occur in molecular graphene \cite{Juan} by displacements of the CO molecules. Specifically, it is possible to show that while massive Dirac fermions occur whenever the renormalization of the on-site energies of the carbon atoms sitting on top of the In and Te sites are equal in sign, gapless behavior with a QBCP is expected in the opposite regime of on-site energies renormalizations opposite in sign. 
The transition from the gapless to the gapped regime is then marked by the emergence of a pseudo-spin one Dirac cone, and can only occur if the effect of the indium layers is negligible as compared to the one due to the chalcogen layers.  
§
In conclusion we have investigated the electronic properties of single layer graphene on top of In$_{2}$Te$_{2}$ monolayer by means of density-functional theory calculations using LDA and HSE. 
In a perfectly commensurate state, this heterostructure can realize two possible inequivalent geometrical configuration. The first configuration, where the In and Te atoms lie above the center of the graphene hexagons, reveals a gap of 20 meV resulting from the formation of a Kekul\'e distortion in the carbon-carbon bonds entirely due to the graphene-indium chalcogenide interaction. This finding suggests that  by experimentally creating vortices of the Kekul\'e texture at a Y junction, a route towards fractionally charged topological excitations -- mathematical analogs of fractional vortices in $p$-wave superconductors -- will be opened. The alternative, almost degenerate, geometrical configuration where the In and Te atoms are above the carbon atoms of graphene, has instead a gapless spectrum with 
the appearance of either an undoubled pseudo-spin one Dirac cone or a quadratic band crossing point at the Fermi level, a bandstructure which is prone to many-body instabilities towards topological Mott insulating states.

GG and MC are financed by European Research Council under FP7/ERC Starting Independent Research Grant ``SUPERBAD" (Grant Agreement n. 240524). 
CO acknowledges the financial support of the Future and Emerging Technologies (FET) programme within the Seventh Framework Programme for Research of the European Commission, under FET-Open grant number: 618083 (CNTQC).


\begin{thebibliography}{99}
\bibitem{EsakiTsu} L. Esaki and R. Tsu, IBM J. Res. Dev. 14, 61 (1970).

\bibitem{Albrecht} C. Albrecht, J.~H. Smet, D. Weiss, K. von.~Klitzing, R. Hennig, M. Langenbuch, M. Suhrke, U. R\"ossler, V. Umansky, and H. Schweizer, Phys. Rev. Lett. {\bf 83}, 2234 (1999). 

\bibitem{Geim1} A. K. Geim and K. S. Novoselov, Nat. Mater. {\bf 6}, 183 (2007).

\bibitem{GiovannettiGRBN} G. Giovannetti, P.~A. Khomyakov, G. Brocks, P.~J. Kelly, and J. van den Brink, Phys. Rev. B {\bf 76}, 073103 (2007).

\bibitem{OrtixInc} C. Ortix, L. Yang, J. van den Brink, Phys. Rev. B {\bf 86}, 081405(R) (2012).

\bibitem{Wallbank} J.~R. Wallbank, A.~A. Patel, M. Mucha-Kruczynski, and V. I. Fal'ko, Phys. Rev. B {\bf 87}, 245408 (2013).

\bibitem{Yankowitz} M. Yankowitz, J. Xue, D. Cormode, J.~D. Sanchez-Yamagishi, K. Watanabe, T. Taniguchi, P. Jarillo-Herrero, P. Jacquod, and B.~J. LeRoy, Nat. Phys. {\bf 8}, 382 (2012). 

\bibitem{Geim2} L. A. Ponomarenko, R. V. Gorbachev, G.~L. Yu, D.~C. Elias, R. Jalil, A.~A. Patel, A. Mishchenko, A.~S. Mayorov, C.~R. Woods, J.~R. Wallbank, M. Mucha-Kruczynski,	 B.~A. Piot, M. Potemski, I.~V. Grigorieva, K.~S. Novoselov, F. Guinea, V.~I. Fal'ko, and A.~K. Geim, Nature {\bf 497}, 594  (2013). 

\bibitem{KIM} C. R. Dean,	 L. Wang,	 P. Maher,	 C. Forsythe, F. Ghahari, Y. Gao, J. Katoch, M. Ishigami, P. Moon, M. Koshino, T. Taniguchi, K. Watanabe, K.~L. Shepard, J. Hone, and P. Kim,  Nature { \bf 497}, 598 (2013).

\bibitem{Hunt} B. Hunt, J.~D. Sanchez-Yamagishi, A.~F. Young, M. Yankowitz, B.~J. LeRoy, K. Watanabe, T. Taniguchi, P. Moon, M. Koshino, P. Jarillo-Herrero, and R.~C. Ashoori, Science {\bf 340}, 1427 (2013). 

\bibitem{Hofstadter} D.~R. Hofstadter, Phys. Rev. B {\bf 14},  2239  (1976).

\bibitem{Woods} C.~R. Woods, L. Britnell, A. Eckmann, R.~S. Ma, J.~C. Lu, H.~M. Guo, X. Lin, G.~L. Yu, Y. Cao, R.~V. Gorbachev, A.~V. Kretinin, J. Park, L.~A. Ponomarenko, M.~I. Katsnelson, Yu.~N. Gornostyrev, K. Watanabe, T. Taniguchi, C. Casiraghi, H.~-J. Gao, A.~K. Geim, K.~S. Novoselov, Nat. Phys. {\bf 10}, 451 (2014).


\bibitem{Falko} J. R. Wallbank, M. Mucha-Kruczynski, and V. I. Fal'ko, Phys. Rev. B {\bf 88}, 155415 (2013).

\bibitem{Falko2}  V. Z\'olyomi, N.~D. Drummond, and V.~I. Fal'ko, Phys. Rev. B {\bf 89}, 205416 (2014). 


\bibitem{Falko3} V. Z\'olyomi, N.~D. Drummond, and V.~I. Fal'ko, Phys. Rev. B {\bf 87}, 195403 (2013). 

\bibitem{Gomes} K.~K. Gomes,	 W. Mar, W. Ko,	 F. Guinea, and  H.~C. Manoharan, Nature {\bf 483}, 306 (2012).

\bibitem{Hou} C.~-Y. Hou, C. Chamon, and C. Mudry, Phys. Rev. Lett. {\bf 98}, 186809 (2007). 

\bibitem{Sera} B. Seradjeh, and M. Franz, Phys. Rev. Lett. {\bf 101}, 146401 (2008). 

\bibitem{Ryu} S. Ryu, C. Mudry, C.~-Y. Hou, and C. Chamon, Phys. Rev. B {\bf 80}, 205319 (2009). 

\bibitem{Green} D. Green, L. Santos, and C. Chamon, Phys. Rev. B {\bf 82}, 075104 (2010).

\bibitem{OrtixArX} M. Manzardo, J. W. F. Venderbos, J. van den Brink, C. Ortix,  arXiv:1402.3145.


\bibitem{DFT} P. Hohenberg and W. Hohn, Phys. Rev. {\bf 136} B864 (1964); W. Kohn and L. J. Sham, Phys. Rev. {\bf 140} A1133 (1965).

\bibitem{VASP} G. Kresse and J. Furthmuller, Phys. Rev. B {\bf 54}, 11169 (1996); Comput. Mater. Sci. {\bf 6}, 15 (1996).

\bibitem{DipoleCorr} J. Neugebauer and M. Scheffler, Phys. Rev. B {\bf 46}, 16067 (1992).

\bibitem{QE} P. Giannozzi, S. Baroni, N. Bonini, M. Calandra, R. Car, C. Cavazzoni, D. Ceresoli, G. L. Chiarotti, M. Cococcioni, I. Dabo, A. Dal Corso, S. Fabris, G. Fratesi, S. de Gironcoli, R. Gebauer, U. Gerstmann, C. Gougoussis, A. Kokalj, M. Lazzeri, L. Martin-Samos, N. Marzari, F. Mauri, R. Mazzarello, S. Paolini, A. Pasquarello, L. Paulatto, C. Sbraccia, S. Scandolo, G. Sclauzero, A. P. Seitsonen, A. Smogunov, P. Umari, R. M. Wentzcovitch, J. Phys. Condens. Matter {\bf 21}, 395502 (2009).

\bibitem{LDA} J. P. Perdew and A. Zunger, Phys. Rev. B {\bf 23}, 5048 (1981).

\bibitem{GiovannettiDopingGR} G. Giovannetti, P. A. Khomyakov, G. Brocks, V. M. Karpan, J. van den Brink, and P. J. Kelly, Phys. Rev. Lett. {\bf 101}, 026803 (2008)

\bibitem{Menno} M. Bokdam, T. Amlaki, G. Brocks, and P. J. Kelly, Phys. Rev. B {\bf 89}, 201404(R) (2014).

\bibitem{Sachs} B. Sachs, T. O. Wehling, M. I. Katsnelson, and A. I. Lichtenstein, Phys. Rev. B {\bf 84}, 195414 (2011)

\bibitem{HSE} J.Heyd, G. E. Scuseria, and M. Ernzerhof, J. Chem. Phys. {\bf 118}, 8207 (2003); J. Heyd and G. E. Scuseria, J. Chem. Phys. 124, 219906 (2006)

\bibitem{Fuchs} F. Fuchs, J. Furthm\"uller, F. Bechstedt, M. Shishkin, and G. Kresse, Phys. Rev. B {\bf 76}, 115109 (2007).

\bibitem{Bergam} D. L. Bergman, Phys. Rev. B {\bf 87}, 035422 (2013).

\bibitem{Sun} K. Sun, H. Yao, E. Fradkin, and S.~A. Kivelson, Phys. Rev. Lett. {\bf 103}, 046811 (2009). 

\bibitem{Dora} B. D\'ora, I.~F. Herbut, and R. Moessner, Phys. Rev. B {\bf 90}, 045310 (2014). 

\bibitem{Juan} F. de Juan, Phys. Rev. B {\bf 87}, 125419 (2013).

\end{thebibliography}
\end{document}